\documentclass[iop]{emulateapj}

\usepackage[applemac]{inputenc}
\usepackage[T1]{fontenc}
\usepackage[english]{babel}
\usepackage{graphicx}

\shorttitle{Constant V and B in Alfvénic fluctuations}
\shortauthors{Matteini et al.} 

\begin{document}

\title{Ion kinetic energy conservation and magnetic field strength constancy in multi-fluid solar wind Alfv\'enic turbulence}

\author{L. Matteini\altaffilmark{1}, T. S. Horbury\altaffilmark{1}, F. Pantellini\altaffilmark{2}, M. Velli\altaffilmark{3}, and S. J. Schwartz\altaffilmark{1}}
\altaffiltext{1}{The Blackett Laboratory, Imperial College London, SW7 2AZ, UK}
\altaffiltext{2}{LESIA, Observatoire de Paris, CNRS, UPMC, Université Paris-Diderot, 
5 place Jules Janssen, 92195 Meudon, France}
\altaffiltext{3}{Department of Earth, Planetary, and Space Sciences, UCLA, California, US}

\date{\today}

\begin{abstract}

We investigate properties of the plasma fluid motion in the large amplitude low frequency fluctuations of highly Alfvénic fast solar wind.
We show that protons locally conserve total kinetic energy when observed from an effective frame of reference comoving with the fluctuations.
For typical properties of the fast wind, this frame can be reasonably identified by alpha particles, which, owing to their drift with respect to protons at about the Alfvén speed along the magnetic field, do not partake in the fluid low frequency fluctuations.
Using their velocity to transform proton velocity into the frame of Alfvénic turbulence, we demonstrate that the resulting plasma motion is characterized by a constant absolute value of the velocity, zero electric fields, and aligned velocity and magnetic field vectors as expected for unidirectional Alfvénic fluctuations in equilibrium.
We propose that this constraint, via the correlation between velocity and magnetic field in Alfv\'enic turbulence, is at the origin of the observed constancy of the magnetic field: while the constant velocity corresponding to constant energy can be only observed in the frame of the fluctuations, the correspondingly constant total magnetic field, invariant for Galilean transformations, remains the observational signature, in the spacecraft frame, of the constant total energy in the Alfv\'en turbulence frame. 
\end{abstract}


\maketitle

\section{Introduction}
The solar wind constitutes a unique laboratory for plasma turbulence \citep[][]{Bruno_Carbone_2013}.
Velocity and magnetic field fluctuations, especially in the fast wind, are known to be Alfvénic \citep{Belcher_Davis_1971, Smith_al_1995}, with correlations between velocity and magnetic field compatible with a unidirectional flux of anti-sunward waves. 
The low frequency part of the electromagnetic spectrum, referred to as the 1/f range, is considered as the reservoir of energy for the turbulent cascade that extends down to the small kinetic scales of the plasma, where it is dissipated via wave-particle interactions or other processes that are still not entirely understood \citep{Leamon_al_1999, Alexandrova_al_2009}.
The low frequency magnetic field fluctuations, $\delta\bf{B}$, have large amplitudes which are often of the same order as the underlying magnetic field, $\bf{B}_0$. 
As a consequence the angle between the direction of the local magnetic field, $\bf{B}=\bf{B}_0+\delta\bf{B}$, and the radial direction, corresponding to the direction of the flow, is observed to fluctuate significantly.
\begin{figure}[h]
\includegraphics[width=8.5cm]{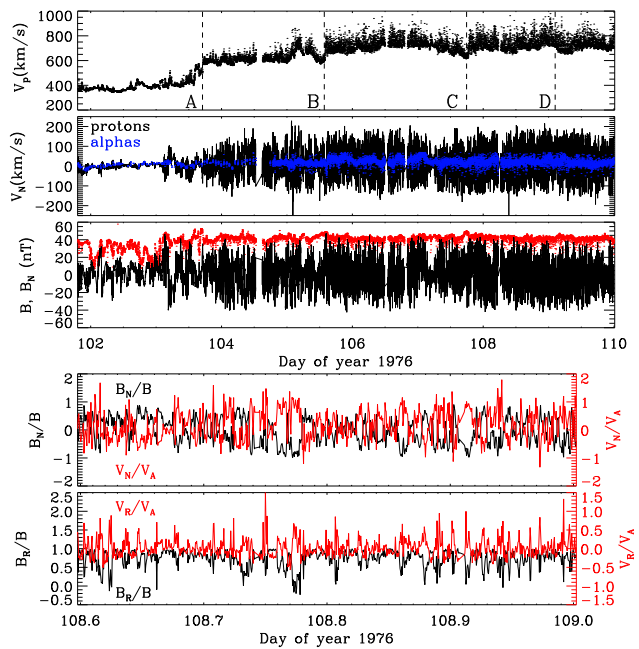}
\caption{Top 3 panels: data used in this work: proton speed (top), normal component of proton velocity (middle) and magnetic field (bottom), and total magnetic field $B$ (red, bottom panel). 
Lower 2 panels: normal (top) and radial (bottom) components of the proton velocity (red) and magnetic field (black), for a $8$hrs interval selected between labels $\rm{A}$ and $\rm{B}$.\label{fig1}}  
\end{figure}

Despite the large excursions of the magnetic field vector, the total magnetic field intensity $|\bf{B}|$ displays a much smaller variance and is observed to remain remarkably constant during highly Alfvénic periods, regardless of heliocentric distance and latitude \citep{Bavassano_Smith_1986, Smith_al_1995}. Geometrically, this means that the tip of the total magnetic field vector $\bf{B}$ moves on a sphere of constant radius \citep{Barnes_1981, Bruno_al_2001} and that fluctuations can not be described by simple planar waves \citep{Goldstein_al_1974, Barnes_1976, Webb_al_2010}. The origin of this remarkable property is still an open question. Large amplitude purely transverse waves propagating in one direction with total field $B=\rm{const}$ are an exact solution of  the MHD equations \citep[e.g.,][]{Barnes_Hollweg_1974}, suggesting that the solar wind plasma is in equilibrium with an ensemble low frequency Alfv\'enic fluctuations propagating away from the sun, but how this condition is achieved in the turbulent expanding solar wind and which dynamical driver is leading to it are not well understood.
Moreover, in the solar wind, ion species (protons and alpha particles) are observed to interact differently with the fluctuations, as a function of their relative drift speed \citep{Goldstein_al_1996}.

In this work we demonstrate, using in situ spacecraft observations and focussing on the 3-D nature of the low frequency Alfvénic fluctuations and the multi-species composition of the solar wind plasma, that the property of constant magnetic field is related to a more fundamental physical property, 
namely the local conservation of the ensemble particle kinetic energy in the effective wave frame of reference.

\section{Data Analysis}
\subsection{Properties of highly Alfvénic fast streams}
Figure \ref{fig1} shows the solar wind data, measured by the Helios spacecraft at 0.3 AU, used in this work (Helios 2, 1976 days 102-110, time resolution 40s).
The top panel shows the proton speed; this includes a very high speed stream \citep{Marsch_al_1982a}. The predicted Parker spiral angle for such conditions is $\sim10$ degrees, so that the average magnetic field should be directed close to the radial direction $R$.
The second and third panels show the $N$ component of the velocity and magnetic field, $V_N$ and $B_N$, perpendicular to the plane that contains both the radial and ${\bf B}_0$. The $T$ direction completes the orthogonal $RTN$ spacecraft coordinate system used in this work.
As the spacecraft enters the fast wind stream an increase in amplitude of the fluctuations is observed. This also corresponds to an increase of Alfvénicity, or correlation between velocity and magnetic field fluctuations \citep{Bruno_al_1985}. However, the amplitude of the velocity fluctuations for alphas  (blue line in the second panel) remains small with respect to protons.
As mentioned, despite the fact that the fluctuating magnetic field shows variation of the same order as the background field ($|\delta {\bf B}|/B\sim1$), the total magnetic field $B$ (red points in third panel) remains approximatively constant over the period of the oscillations. 
Note that over the fast wind stream the direction of the average magnetic field remains approximatively constant, while the local instantaneous magnetic field $\bf{B}$ oscillates from 0 to beyond 90 degrees.

The two lower panels show a $\sim10$hr sub-interval, demonstrating the high level of Alfvénicity in the fast stream, i.e. the anti-correlation between magnetic and velocity fluctuations.
The $N$ direction displays roughly symmetric fluctuations.
By contrast, the $R$ component of the fluctuations (bottom panel) is one-sided, as already discussed in \cite{Gosling_al_2009}; this can be explained in terms of the geometry of Alfvénic fluctuations with constant magnetic field \citep{Matteini_al_2014}.

\begin{figure}
\includegraphics[width=8.5cm]{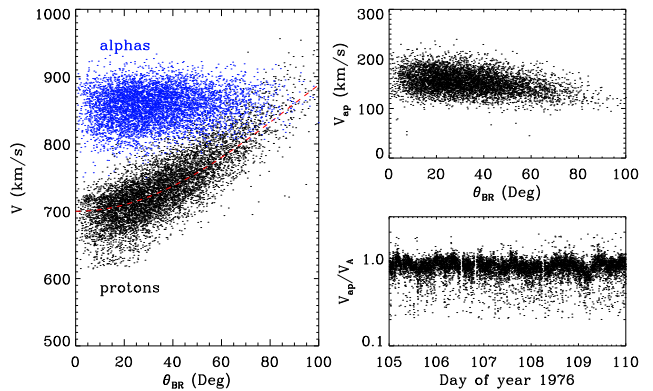}
\caption{Left: Proton and alpha (blue) speed as a function of $\theta_{BR}$, angle between the radial and the local magnetic field $\bf{B}$. Dashed red line indicates the correlation discussed in the text \citep{Matteini_al_2014}. Right: Alpha-proton drift $V_{\alpha p}$ as a function of $\theta_{BR}$ (top) and in units of the local Alfvén speed $V_A$ (bottom).}
\label{fig2} 
\end{figure}

Some other typical properties of the highly Alfvénic fast solar wind interval considered here are illustrated in Figure~\ref{fig2}.
The left panel shows the correlation between the proton speed (black dots) and the angle between the radial and the local direction of the magnetic field $\theta_{BR}$ over the interval B-D. As discussed by \cite{Matteini_al_2014} this effect is due to the presence of large amplitude Alfvénic fluctuations; as a consequence the proton speed can be reasonably described by the relation: 
\begin{equation}\label{eq_vp}
V_p\sim V_0+A[1-\rm{cos}(\theta_{BR})], 
\end{equation}
shown in red line, with $V_0=700$ km/s and $A=160$ km/s, and where $V_0$ is the minimum of the average solar wind speed over the interval, 
and $A$ is the "phase" velocity of the fluctuations. The latter is proportional to the Alfvén speed ${V_A: A=V_A\sqrt{r_A}}$ through the Alfvén ratio $r_A$, the ratio between velocity and magnetic energy of the fluctuations. 
In fast streams typically $r_A\le1$, depending on heliocentric distance.
Since $r_A$ decreases with increasing radial distance, the deviation of $A$ from the Alfvén speed becomes more significant for observations far from the Sun, as for example it is found in the Ulysses data set \citep[cfr.][]{Matteini_al_2014}. In Fig.~\ref{fig2} the value of $A$ is quite close to the measured Alfvén speed, due to the small deviation from unity of the Alfvén ratio $r_A$ in this interval \citep{Bruno_al_1985}.

Note that the scaling of Eq. (\ref{eq_vp}) also provides an explanation for the radial one-sided fluctuations on the bottom panel of Fig \ref{fig1}.
Being  
\begin{equation}\label{eq_vp2}
V_p=\sqrt{(V_0+\delta V_R)^2+\delta V_\perp^2}
\end{equation}
and since $\delta V_R\sim \delta V_\perp \ll V_0$, at first order:
\begin{equation}\label{eq_vr}
 V_p\sim V_0+\delta V_R\,.
\end{equation}
From Eq.~(\ref{eq_vp}), we then expect $\delta V_R\propto 1-\rm{cos}(\theta_{BR})$ and since $\theta_{BR}$ roughly oscillates between $[0,\pi/2]$ this leads to a flatter distribution when $\theta_{BR}\sim0$ and spike-like enhancements when $\theta_{BR}\sim90$.

Unlike the protons, the speed of alpha particles $V_\alpha$(blue dots) is not correlated with $\theta_{BR}$. The difference between proton and alpha speed $V_\alpha-V_p$ is therefore a function of the direction of the local magnetic field and $V_\alpha\sim V_p$ when $\theta_{BR}\sim90^{\circ}$. It is also worth underlining that the such a plasma motion implies effective changes in the speed of the centre of mass as a function of $\theta_{BR}$.
However the drift speed between alphas and protons \mbox{$V_{\alpha p}=|\bf{V}_\alpha-{\bf V}_p|$}, which to a good approximation is always aligned with the local magnetic field  \citep{Marsch_al_1982a}, does not change during Alfvénic fluctuations, and is roughly independent of $\theta_{BR}$ (upper right panel). 
The bottom right panel shows the value of $V_{\alpha p}$ normalized to the local Alfvén speed over the period analyzed here. $V_{\alpha p}$ is a significant fraction of $V_A$ and the ratio is close to 1 in the fast stream,  with an average value $\left<V_{\alpha p}\right>\sim0.85V_A$ over the period shown.


All these properties, i.e. the fact that alphas stream faster than protons at about the Alfvén speed and that their drift with respect to protons does not change in time, explain why the alpha particle speed is observed to be uncorrelated with the angle $\theta_{BR}$: since alphas travel at approximatively the phase speed of the turbulence, they do not respond to the oscillations of the magnetic field; alphas surf the large amplitude fluctuations of the solar wind \citep{Marsch_al_1981} and their speed, unlike the protons, is not modulated by the Alfvénic activity \citep{Goldstein_al_1996}.
Similar behavior is observed for the minor heavy ions \citep{Berger_al_2011, Gershman_al_2012} which also stream faster than protons in the fast solar wind. 

\subsection{Particle motion and constant magnetic field}
It is worth noting that not all the characteristics of the solar wind summarized in Figures \ref{fig1} and \ref{fig2} are fully understood.
In particular, despite the fact that fluctuations propagating in a single direction with a total constant magnetic field magnitude are a nonlinear solution to the MHD equations, so that the fluctuations may in some sense be considered to be in equilibrium, how this is achieved in the solar wind plasma is still an open question. For example, though monochromatic Alfvén waves with circular polarization are clearly an example solution with constant total magnetic field \citep[e.g.,][]{Barnes_Hollweg_1974}, a collection of transverse Alfvén waves will not satisfy $|{\bf B}|=\rm{const}$, unless their phases are closely correlated in a very specific way. 
More generally fluctuations with amplitude $|{\bf \delta B}|$  modulated in time, as in Figure~\ref{fig1}, can not fulfill the condition of constant total magnetic field: 
\begin{equation}
|{\bf B}|^2=|{\bf B_0}+{\bf \delta B}|^2=B_0^2+|2{\bf B_0}\cdot{\bf \delta B}|+|{\bf \delta B}|^2={\rm const.}
\end{equation}
if they are 2D lying in the plane transverse to ${\bf B}_0$ \citep[e.g.,][]{Dobrowolny_al_1980a}. On the contrary, magnetic fluctuations in the solar wind are observed to be 3-D (i.e., ${\bf B}_0/B_0\cdot\delta {\bf B}=\delta B_\|\ne0$), with the magnetic field vector moving on a sphere of constant radius. 

In this framework, in order to maintain equilibrium between particles and fluctuating fields, thus avoiding energy exchanges leading to damping, 
one condition is that in the oscillating fields the plasma conserves energy as seen \emph{in the frame moving with the fluctuations}.
The conservation of total kinetic energy in the wave frame would then imply that protons move on a particular surface in phase space.

\begin{figure}
\includegraphics[width=8.5cm]{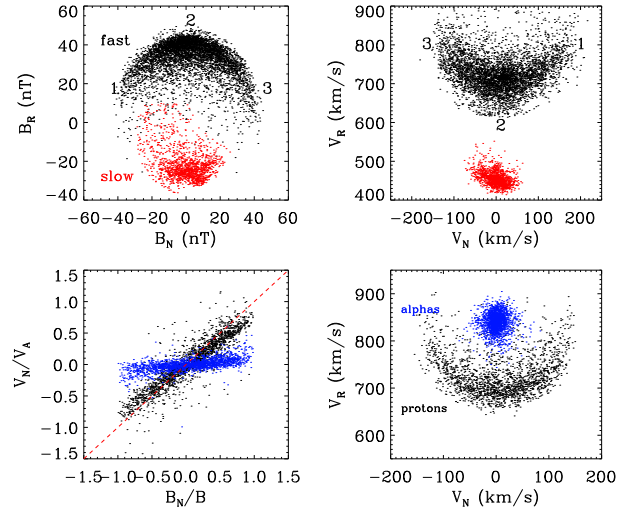}   
\caption{Top: scatterplot of the normal and radial magnetic field (left) and proton velocity (right) components, for the interval B-D in Fig \ref{fig1}. Red points show the preceding slow wind interval as reference (before label A in Fig \ref{fig1}). Numbered labels indicate the corresponding anti-correlation between $B_R$ and $V_R$.
Bottom: (left) scatterplot of normal Alfvénic components for protons and alphas (blue) for the interval C-D. The red dashed line identifies the slope corresponding to the Alfvén speed. (right) scatterplot of proton and alpha (blue) velocity in the plane ($V_N,V_R$) for the same interval as left panel.
\label{fig3}}
\end{figure}

We now show that such a surface exists in the solar wind and can be identified with good accuracy from observations. Figure \ref{fig3}, top left panel, shows the excursion of $\bf{B}$ during the period analyzed, for fast (black) and slow (red) wind, in the $(B_N,B_R)$ plane.
It can be clearly seen that in the fast wind,  the magnetic field oscillates on an arc of constant radius; this is the projection on the plane of the excursion of $\bf{B}$ on a sphere.
The picture is less clear, and the level of the fluctuations is weaker, in the short interval of slow wind preceding the fast stream, reported in red for comparison.
The top right panel shows the analogous evolution of the velocity vector ${\bf V}_p$ in the plane $(V_N,V_R)$:
in the fast wind the arc-shaped excursion in the magnetic field corresponds to oscillations of the velocity on an analogous arc as a function of $V_N$ and $V_R$. It can be shown that the 3D motion of the velocity vector identifies a spherical surface, consistent with the motion of $\bf{B}$.
Note that, due to the Alfvénic anti-correlation between the components of $\bf{B}$ and $\bf{V}$, minima of $B_R$ (labelled as 1 and 3 in the figure) correspond to maxima in $V_R$, and viceversa (labelled as 2 in the figure).

As mentioned, alpha particles do not partake in the Alfvénic motion when they stream close to the Alfvén speed.
The bottom left panel of Fig.~\ref{fig3} displays the correlation between $B_N/B$ and $V_N/V_A$, for both protons and alphas (blue). As discussed by \cite{Goldstein_al_1996}, while protons show a strong correlation between velocity and magnetic field fluctuations, close to the relation expected for fluctuations that propagate at the Alfvén speed (red dashed line), the alpha velocity has only a weak correlation with Alfvénic magnetic fluctuations. The transverse motion of alphas is almost negligible compared to the amplitude of the proton fluctuations; as a consequence, alphas do not trace an arc in the $(V_N,V_R)$ plane (right panel), but rather remain in a fixed place in phase space. The position of alpha particle in velocity space therefore identifies to a good approximation the location of the wave frame, about which the protons oscillate. Although it is not always possible to use alphas in this way to identify the wave frame, the frame may also be found by direct inspection of the geometry of the fluctuations, such as the slope of the $V_N$-$B_N/B$ correlation, which provides an independent estimation of the phase speed  associated to the fluctuations \citep[see for example][for more details]{Goldstein_al_1996}. 
In this case, the phase speed inferred from the $V_N$-$B_N/B$ correlation is $\sim155$ km/s, thus slightly less ($\sim0.9V_A$) than the mean Alfvén speed measured during this interval, $\left<V_A\right>\sim175$km/s. 
We note here that, remarkably, the mean alpha-proton drift shown in the right-bottom panel of Fig.~\ref{fig2} is also close to this value ($\left<V_{\alpha p}\right>\sim0.85V_A$), suggesting that alphas are drifting with respect to protons with a speed which is closer to the phase speed inferred from the analysis of the plasma motion rather than the nominal Alfvén speed. This then confirms that alphas are essentially comoving with the fluctuations, and are thus at rest with them, consistently with the fact that they do not display velocity oscillations in the bottom panels of Fig.~\ref{fig3}.

\begin{figure}
\includegraphics[width=9cm]{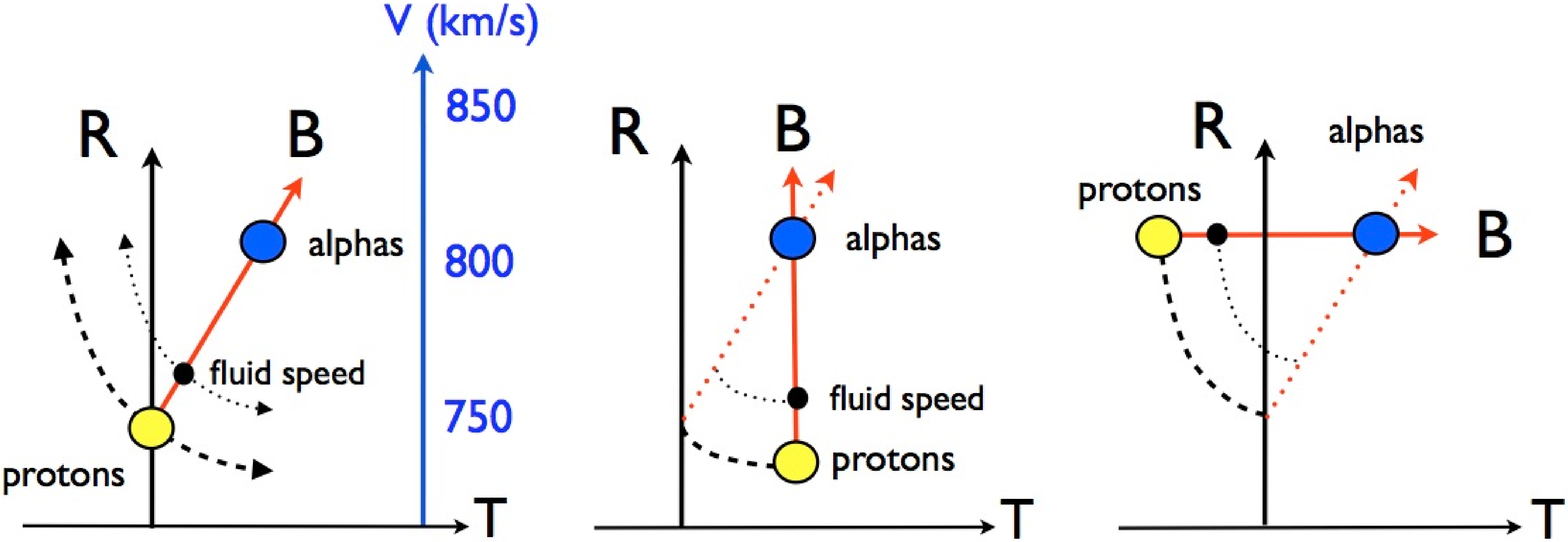}
\caption{Schematic description of plasma motion in large amplitude low frequency Alfvénic fluctuations in the velocity plane ($V_T,V_R$). Circles identify protons (yellow), alphas (blue), and the fluid (centre of mass) frame (black). Middle and right panels show the case of a radially aligned and transverse local magnetic field (red arrow), respectively.}  
\label{fig4}
\end{figure}

The geometry of this dynamics is summarized by the cartoon of Figure~\ref{fig4}. The left panel shows a schematic representation of the solar wind plasma: protons (yellow circle) have a purely radial velocity ($\sim750$ km/s) and alphas stream faster with respect to them, along the local magnetic field. The mean magnetic field is assumed to have a small angle in the $R$-$T$ plane, reproducing the Parker spiral angle at 0.3 AU. The small black circle identifies the fluid velocity ($V_f={{N_pV_p+4N_\alpha V_\alpha}\over{N_p+4N_\alpha}}$), which lies close to the proton frame. Oscillations of the proton velocity imposed by the low frequency Alfvénic turbulence cause correlated changes in the proton speed and magnetic field vector, while they do not change significantly the alpha speed (See Figure \ref{fig2}).
When the local magnetic field rotates towards large $\theta_{BR}$ (right panel), protons are observed to speed up \citep{Matteini_al_2014}, and $V_p\sim V_\alpha$.
When the magnetic fluctuations are such that $\bf{B}$ becomes more radial, protons decelerate a bit and the difference between proton and alpha speeds is maximum (middle panel). 

\begin{figure*}
\includegraphics[width=17.4cm]{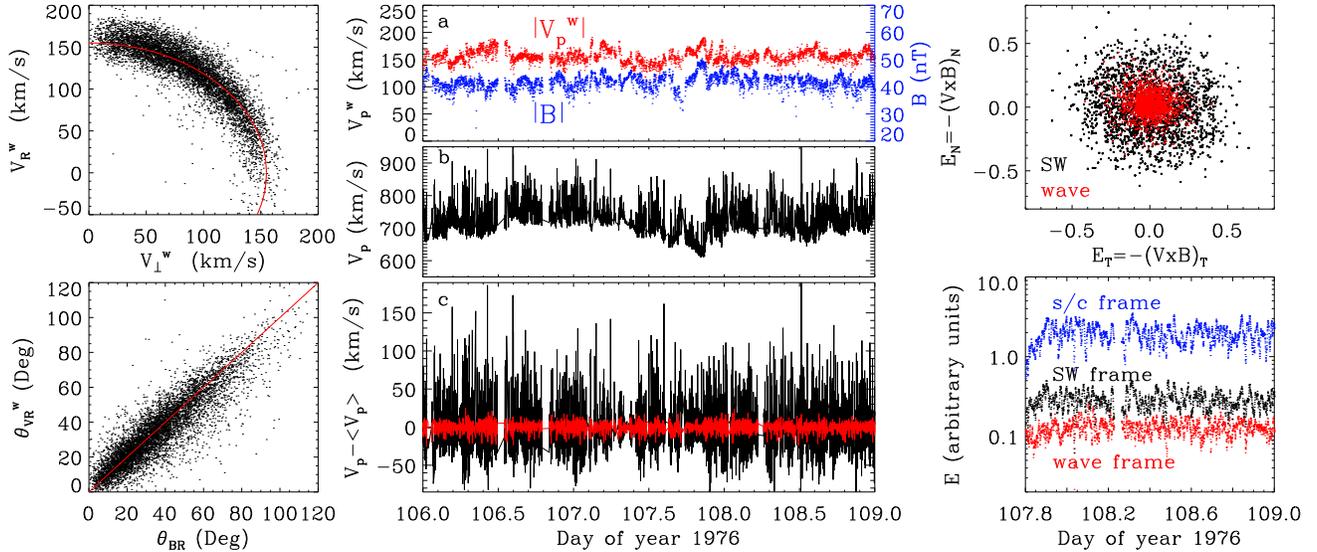}
\caption{ Left panels: Proton velocity in the estimated frame of Alfvénic fluctuations; (top) velocity components in $(V_\perp^{w},V_R^{w})$, red line corresponds to the condition $|\bf{V}|=\rm{const}$; (bottom) correlation between the angle of the magnetic field, $\theta_{BR}$, and of the proton velocity,  $\theta_{VR}^w$; the red line corresponds to ${\bf V}=\alpha \bf{B}$. Middle panels: proton speed $V_p$ in (a) wave (red) and (b) spacecraft frames and (c) associated variations; blue dots in the top panel refer the $|\bf{B}|$ for comparison. Right: (top) components of the motional electric field $\bf{E}=-\bf{V}\times \bf{B}$ estimated in the plasma and wave (red) frames. (bottom) Comparison between $|\bf{E}|$ estimated in spacecraft (blue), plasma (black) and wave (red) frames.}  
\label{fig5}
\end{figure*}
\subsection{Plasma properties in the frame of fluctuations}
Keeping in mind the picture of Figure~\ref{fig4} and what we have discussed so far, we can now conclude our analysis.
The shape of proton velocity in phase space shown in Figure~\ref{fig3} suggests the presence of a particular frame that corresponds to the pivot of the proton oscillation, which can be to a first approximation identified by the velocity of the alphas in the same phase space; such a frame is obviously the wave frame\footnote{This is true because to the properties of the high stream selected and generally common in the fast wind. However, the same description will hold in the absence of a proton-alpha drift; we know that in such a case alphas partake in wave motion as protons. In that situation, the wave frame will obviously not be identified by any particular species and should be estimated through particle Alfvénic motion.} 
In this frame the electric field associated with the fluctuations is expected to vanish\footnote{This is valid within ideal MHD when contribution of the Hall term is negligible, i.e. as long as the difference between the ion frame and the electron (plasma) frame is considered small, a condition satisfied by the low frequency fluctuations investigated here.}: 
\begin{equation}\label{eq_1}
{\bf{E}}=-({\bf{V}}\times{\bf{B}})=0 \,, 
\end{equation}
meaning also that each species $i$ maintains its total velocity ${\bf V}_i$ aligned with the instantaneous magnetic field, i.e.:
\begin{equation}\label{eq_2}
{\bf V}_i=\alpha_i {\bf B} \,; \ \ \ \ \  |{\bf V}_i|=|\alpha_i {\bf B}|=\rm{const.}
\end{equation}
Note that such conditions are expected for non-dispersive monochromatic transverse waves with circular polarization in a multi-fluid plasma \citep{Marsch_Verscharen_2011}. The constant of proportionality $\alpha$ depends on each species' drift with respect to the wave frame, ${\alpha_i=V_{0i}/B_0}$.
Here, for alphas $\alpha\sim0$, while for protons ${\alpha\sim V_a/B_0}$.
Therefore particles of each species move on a surface of constant energy as discussed previously; these conditions can be directly tested on the observations, transforming velocities measured by the spacecraft into the estimated wave frame.

Figure~\ref{fig5} shows the results of such a transformation where the superscript $w$ indicates velocities in the new (wave) frame. Panels on the lefthand side show (top) the proton velocity in the $(V_\perp^{w},V_R^{w})$ plane, with $V_\perp=\sqrt{V_N^2+V_T^2}$ and (bottom) the correlation between the angle, with respect to the radial, of the instantaneous magnetic field vector, $\theta_{BR}$, and of the proton velocity, $\theta_{VR}^w$. 
These are computed using the alpha velocity for the coordinate change, over the interval B-D of Figure~\ref{fig1}. 
Note that this has the effect of removing the solar wind underlying structures \citep{Thieme_al_1989}, isolating the contribution of Alfvénic fluctuations.
Both panels show that the motion of protons as seen in the new frame well satisfies Eq.~(\ref{eq_2}), corresponding to a good approximation to a motion with the proton velocity aligned with the instantaneous magnetic field and on a sphere of constant radius, so that $|{\bf{V}}_p|$, and hence the proton kinetic energy, are constant. The radius of the constant energy sphere corresponds to the alpha-proton drift speed and is consistent, as discussed previously, with the phase speed of the fluctuations that can be obtained from the correlation in the left bottom panel of Figure~\ref{fig3} over the whole interval.
The middle ($\tt{a}$) panel of Figure~\ref{fig5} shows in red the temporal variation of the new proton speed $V_p^w$: remarkably, it remains roughly constant over the whole duration of the high speed stream, consistently with the variation of the total magnetic field $|\bf{B}|$ (shown in blue).
The proton speed profile $V_p$  measured in the spacecraft frame is reported for comparison in the ($\tt{b}$) panel. This is characterized by large scale modulation (microstreams lasting on the order of a day \citep{Neugebauer_al_1995}) and by Alfvénic oscillations on the minute scale. 
As expected, the latter fluctuations are larger in $V_p$ than in $V_p^w$.
This is further demonstrated by the middle ($\tt{c}$) panel, reporting the variations of the proton speed $V_p-\left<V_p\right>$, as measured both in the spacecraft (black) and the wave frame (red). The average $\left<V_p\right>$ is taken over 20 minutes in order to remove the effects of larger scale structures. The fluctuations of the proton speed are minimized when transforming to the wave frame; in this frame the fluctuations become symmetric, removing the one-sided effect observed in the spacecraft frame.

\subsection{Motional electric field: ${\bf E}=-{\bf V}\times {\bf B}$}
Finally, let us consider the electric field associated with the fluctuations.  Despite the 3D nature of the variations in $\bf{B}$ and $\bf{V}$ shown in Figure~\ref{fig3}, the associated electric field, in the fluid frame, is essentially 2D.  This is because the component parallel to the mean magnetic field, 
\begin{equation}
E_\|=-({\bf V}\times \delta{\bf B})\cdot{\bf B}_0/B_0
\end{equation}
essentially vanishes: since the electric field in the wave frame is close to zero, the Lorentz transformation backwards  along the main field, connecting the solar wind plasma frame to the wave frame, does not change the component along ${\bf B}_0$ ($\bf{B}\cdot\bf{E}$ is Lorentz invariant), so $E_\|\sim0$ also in plasma frame. Data confirm that $E_\|$, as computed from Eq. (\ref{eq_1})  and approximatively radial, is significantly smaller than the other electric field components, even though it is not exactly zero due to the level of uncertainties contained in the measurements. 

The right top panel of Figure~\ref{fig5} displays the other components $E_T$ and  $E_N$, computed from Eq.~(\ref{eq_1}), in the average solar wind (proton) frame $\left<V_p\right>$ (black) and in the wave frame (red). As expected the intensity of the electric field is significantly smaller when computed in the wave frame; in that frame $|\bf{E}|$ should ideally vanish, while here the transverse component of $\bf{E}$ approximatively reduces to the amplitude of $E_\|$.
The right bottom panel shows a comparison of $|\bf{E}|$ as computed in different frames, for the interval $C$-$D$  in Figure~\ref{fig1}.
The electric field in the spacecraft frame (blue) contains the contribution of the solar wind motion; this can be removed transforming to the average plasma (proton) frame (black), but still retaining the fluctuating component of $\bf{E}$ due to the Alfvénic motion of protons. Finally, when transforming to the wave frame (red), the latter is removed, and the residual electric field is minimum.
It can be shown that, within observational uncertainties about the speed of the wave frame, such an electric field is the minimum possible \citep[e.g.,][]{Khrabrov_Sonnerup_1998}, since transforming to a different frame along ${\bf B}_0$ produces a larger motional electric field, while a different transformation, with a component orthogonal to ${\bf B}_0$, introduces a component $E_\|\ne0$. 

\section{Conclusion}
In summary, we have shown that the motion of protons in large amplitude low frequency Alfvénic fluctuations in the fast solar wind conserves kinetic energy in the wave frame. Though this might seem a natural consequence of the equilibrium between particles and fluctuations in the unidirectional Alfv\'en-like fluctuations propagating away from the sun, it is quite remarkable that this can be recovered, using in situ observations, in the complex and turbulent dynamics of the solar wind plasma, with fluctuations of the order or larger than 100 km/s. Moreover, this highlights the need for identifying a mechanism able to drive and control such a condition, as well as understanding whether this condition is set already close to the Sun or if it develops dynamically in the expanding solar wind. 

Note that a similar description applies also to components of the plasma which have not been considered here, like secondary proton beams that are  ubiquitously observed in the fast solar wind \citep[e.g.,][]{Marsch_al_1982a, Matteini_al_2013}.
In particular, we expect that beams with a drift speed which is larger than the phase speed of the turbulent fluctuations ($\sim V_A$), would oscillate in anti-phase with protons \citep{Goldstein_al_1996}.
A simplified view of this dynamics is shown in the cartoon of Figure~\ref{fig6}, following Figure~\ref{fig4}, where now the small yellow circle identifies the proton beam population, streaming at $V_b>V_A$ (left panel). 

If a local fluctuation is large enough to reverse the radial component of the magnetic field (right panel), the proton core-beam structure is then reversed as well. This sketch reproduces well the dynamics
observed during magnetic switchbacks in the solar wind \citep{Neugebauer_Goldstein_2012}; in particular it explains how during magnetic field reversals core and beam speeds are observed to flip and why the rotation is not centered on the speed of the center of mass, but rather the speed of alpha particles ($V_{\alpha}\sim V_p+V_A$), resulting then in a significant increase of the plasma bulk speed \citep[see figure 2 of][]{Neugebauer_Goldstein_2012}.

A similar description would also apply to the so-called electron strahl population, which carries most of the solar wind heat flux and follows at good approximation the orientation of the local magnetic field \citep{Feldman_al_1975}.
We then expect that during Alfvénic oscillations electrons from the strahl conserve their kinetic energy in the wave frame of fluctuations, making a rotation in velocity space similarly to the proton beam of Figure \ref{fig6}, but with a much larger radius, owing their large drift (typically $\sim1000 \rm{km/s} \gg V_A$). Confirmation of these expectations will be subject of more detailed future studies

Finally, our findings also lead to the following interpretation about the relative constancy of the magnetic field intensity as ubiquitously measured in highly Alfvénic solar wind streams: as discussed, under the effects of the turbulence, the plasma (protons) undergo large amplitude collective oscillations. In the wave frame, the associated electric field is zero and particles move on a surface of constant energy. This condition on particle velocity translates, due to the high level of Alfvénicity, i.e. correlation or anti-correlation of magnetic field and velocity field fluctuations, into magnetic field fluctuations with the same characteristics, leading to $\bf{V}=\alpha\bf{B}$, with $\alpha$ constant in time; it follows that the tip of the total magnetic field vector also fluctuates on a sphere. When measuring solar wind fluctuations in the spacecraft frame, the condition on particle velocity is not observed  directly because kinetic energy is not invariant in the transformation, leading rather to the  solar wind speed profile of Fig \ref{fig1}. On the other hand the magnetic field, which is invariant for Galilean transformations, maintains the signature $|\bf{B}|=\rm{const}$, as observed.

Note that the data used in this work, measurements of the fast wind at 0.3 AU, are appropriate to emphasize these effects, since the amplitude of velocity fluctuations scales, according to WKB, approximatively as 
\begin{equation}
\delta V\propto \delta B/B\cdot V_A=R^{0.5}\cdot R^{-1}=R^{-0.5}
\end{equation}
and is larger closer to the Sun; however we have checked that the same behavior is observed in other high speed streams at various distances. Further confirmation of our findings will be possible  thanks to the forthcoming missions, Solar Orbiter and Solar Probe Plus, that will explore the internal heliosphere.  
Our results also imply that Alfv\'enic turbulence could lead to an enhancement of radial velocity fluctuations, up to or of the order of the average wind speed, as the absolute maximum of fluctuation amplitude is reached inside or around the Alfv\'en radius, where the solar wind accelerates beyond the Alfv\'en speed.

\begin{figure}[t]
\includegraphics[width=8.5cm]{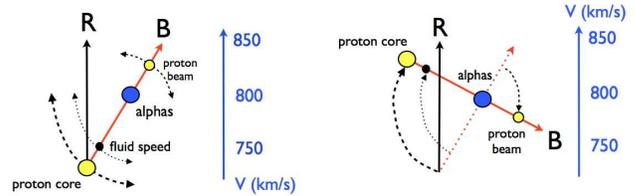}
\caption{Schematic description of plasma motion in Alfvénic fluctuations as in Figure~\ref{fig4}, for distinct proton core and beam populations (large and small yellow circles, respectively). Left panel shows that, since the proton beam is drifting at a speed larger than $V_A$, then it oscillates in anti-phase with the core. Right panel shows that in the case of particularly large amplitude fluctuations producing a reversal of the magnetic field, the core-beam structure is also flipped with respect to the alpha particles frame.}  
\label{fig6}
\end{figure}

\begin{acknowledgments}
Authors acknowledge very stimulating and helpful discussions with S. Landi, A. Verdini, P. Hellinger, R. Grappin, B. Goldstein, and M. Neugebauer.
The research leading to these results has received funding from the UK Science and Technology Facilities Council grant ST/K001051/1 and the European Commission’s Seventh Framework Programme (FP7/2007-2013) under the grant agreement SHOCK (project number 284515).
\end{acknowledgments}

\bibliographystyle{apj.bst}


\end{document}